\documentclass[12pt,letterpaper]{article}

\usepackage{epsfig}

\begin{document}


\begin{center}
{\LARGE\bf Precision study of the $SU(3)$\\[2mm]
topological susceptibility in the continuum}
\end{center}
\vspace{10pt}

\begin{center}
{\large\bf Stephan D\"urr$\,^{a}$}\hspace{2pt}{\large ,}\hspace{4pt}
{\large\bf Zoltan Fodor$\,^{b,c,d}$}\hspace{2pt}{\large ,}\\[1mm]
{\large\bf Christian Hoelbling$\,^{b}$}\hspace{4pt}{\large and}\hspace{4pt}
{\large\bf Thorsten Kurth$\,^{b}$}
\\[10pt]
${}^a\,${\sl Universit\"at Bern, Institut f\"ur theoretische Physik,
Sidlerstr.\,5, CH-3012 Bern, Switzerland}\\
${}^b\,${\sl Bergische Universit\"at Wuppertal,
Gaussstr.\,20, D-42119 Wuppertal, Germany}\\
${}^c\,${\sl E\"otv\"os University, Physics Department, P\'azm\'any 1,
H-1117 Budapest, Hungary}\\
${}^d\,${\sl University of California at San Diego,
9500 Gilman Drive, La Jolla, CA 92093-0319, USA}
\end{center}
\vspace{10pt}

\begin{abstract}
\noindent
We determine the topological susceptibility in the $SU(3)$ pure gauge theory.
We perform a series of high-statistics lattice studies and take the combined
continuum and infinite volume limit. We find
$\chi_\mathrm{top}r_0^4=0.0524(7)(6)$,
which translates into $\chi_\mathrm{top}^{1/4}=193(1)(8)\,\mathrm{MeV}$
with the second error exclusively due to the intrinsic scale ambiguity.
\end{abstract}
\vspace{10pt}


\newcommand{\pad}{\partial}
\newcommand{\psl}{\partial\!\!\!/}
\newcommand{\hqu}{\hbar}
\newcommand{\ovr}{\over}
\newcommand{\til}{\tilde}
\newcommand{\pri}{^\prime}
\renewcommand{\dag}{^\dagger}
\newcommand{\<}{\langle}
\renewcommand{\>}{\rangle}
\newcommand{\gaf}{\gamma_5}
\newcommand{\lap}{\triangle}
\newcommand{\dal}{{\sqcap\!\!\!\!\sqcup}}
\newcommand{\trc}{\mathrm{tr}}
\newcommand{\Mpi}{M_\pi}
\newcommand{\Fpi}{F_\pi}

\newcommand{\al}{\alpha}
\newcommand{\be}{\beta}
\newcommand{\ga}{\gamma}
\newcommand{\de}{\delta}
\newcommand{\ep}{\epsilon}
\newcommand{\ve}{\varepsilon}
\newcommand{\ze}{\zeta}
\newcommand{\et}{\eta}
\renewcommand{\th}{\theta}
\newcommand{\vt}{\vartheta}
\newcommand{\io}{\iota}
\newcommand{\ka}{\kappa}
\newcommand{\la}{\lambda}
\newcommand{\rh}{\rho}
\newcommand{\vr}{\varrho}
\newcommand{\si}{\sigma}
\newcommand{\ta}{\tau}
\newcommand{\ph}{\phi}
\newcommand{\vp}{\varphi}
\newcommand{\ch}{\chi}
\newcommand{\ps}{\psi}
\newcommand{\om}{\omega}

\newcommand{\psb}{\overline{\psi}}
\newcommand{\etb}{\overline{\eta}}
\newcommand{\psd}{\psi^{\dagger}}
\newcommand{\etd}{\eta^{\dagger}}
\newcommand{\qh}{{\hat q}}
\newcommand{\kh}{{\hat k}}

\newcommand{\bdm}{\begin{displaymath}}
\newcommand{\edm}{\end{displaymath}}
\newcommand{\bea}{\begin{eqnarray}}
\newcommand{\eea}{\end{eqnarray}}
\newcommand{\beq}{\begin{equation}}
\newcommand{\eeq}{\end{equation}}

\newcommand{\mr}{\mathrm}
\newcommand{\mb}{\mathbf}
\newcommand{\Nf}{{N_{\!f}}}
\newcommand{\Nc}{{N_{\!c}}}
\newcommand{\ri}{\mr{i}}
\newcommand{\DW}{D_\mr{W}}
\newcommand{\DN}{D_\mr{N}}
\newcommand{\MeV}{\,\mr{MeV}}
\newcommand{\GeV}{\,\mr{GeV}}
\newcommand{\fm}{\,\mr{fm}}
\newcommand{\MSB}{\overline{\mr{MS}}}

\hyphenation{topo-lo-gi-cal simu-la-tion theo-re-ti-cal mini-mum con-tinu-um}


\section{Introduction}


In QCD with $\Nc$ colors and $\Nf$ light dynamical quarks there are two notions
of the topological susceptibility, defined as the second moment of the global
topological charge distribution.

On the one hand, the \emph{actual} topological susceptibility
$\ch_\mr{top}^\mr{QCD}$ shows a clear sensitivity on the dynamical (sea) quark
masses \cite{Leutwyler:1992yt}.
This property renders it an ideal vacuum diagnostics tool, as emphasized in
\cite{Durr:2001ty} and exploited in a number of recent studies
\cite{Bali:2001gk,AliKhan:2001ym,Hart:2001pj,Kovacs:2001bx,Bernard:2003gq,
DeGrand:2005vb,Egri:2005cx,Hasenfratz:2006bq}.

On the other hand, the \emph{quenched} topological susceptibility
$\ch_\mr{top}^\mr{YM}$ is the quantity which is linked via the famous
Witten-Veneziano formula \cite{Witten:1979vv,Veneziano:1979ec}
\beq
\ch_\mr{top}^\mr{YM}\doteq{F^2\ovr2\Nf}(M_{\et'}^2+M_\et^2-2M_K^2)
\label{WV}
\eeq
to the excess of the $\et'$ mass over the pseudoscalar octet masses, with a
proportionality factor which contains the pseudoscalar decay constant%
\footnote{We use the Bern normalization where $\Fpi^\mr{phys}\!=\!92.4(3)\MeV$
and $F\!=\!86.2(5)\MeV$ in the chiral limit.}
in the chiral limit.
This relation is supposed to hold at the leading order in an expansion in
$1/\Nc$ (for a different viewpoint see
\cite{Minkowski:1989vd,Minkowski:1997hv,Girlanda:2001pc}).
To be precise, this quenched susceptibility is the susceptibility of the
underlying $SU(\Nc)$ Yang-Mills (YM) theory, and this is what makes (\ref{WV})
appealing from a theorist's viewpoint -- it relates two different theories.
Specifically, on the lattice one can measure the l.h.s.\ of (\ref{WV}) for
several $\Nc$ and evaluate the r.h.s.\ for various $(\Nc,\Nf)$ combinations,
and finally check whether the agreement is parametrically controlled by $1/\Nc$
or $\Nf/\Nc$.

In this paper we elaborate on the first step in this program -- we determine
with unprecedented precision the quenched topological susceptibility for
the case $\Nc\!=\!3$.
We begin with an exposition of the main avenues towards defining a topological
charge on the lattice and how one extracts the topological susceptibility.
The next two sections contain details of our lattice simulations and of our
combined continuum and infinite volume extrapolation.
Having a result in terms of the Sommer radius $r_0$ \cite{Sommer:1993ce}, the
latter needs to be identified with a length-scale in $\fm$, and we discuss both
the input that goes into such an identification and the remaining ambiguity.
In the concluding section we compare our result to other recent determinations
of $\ch_\mr{top}\equiv\ch_\mr{top}^\mr{YM}$.
Details of a new parameterization of $r_0(\be)$ have been arranged in an
appendix.


\section{Topological charge definition}


In the continuum the topological charge of a given gauge background is
defined as
\beq
q
={1\ovr16\pi^2}\int\!dx\;\trc(F_{\mu\nu}(x)\til F_{\mu\nu}(x))
={1\ovr32\pi^2}\int\!dx\;\trc(\ep_{\mu\nu\si\rh}F_{\mu\nu}(x)F_{\si\rh}(x))
\label{defqcon}
\eeq
where $F_{\mu\nu}\!=\!F_{\mu\nu}^a\la^a/2$ is the field strength tensor.
For toroidal space-time geometry (\ref{defqcon}) is integer and linked to the
index of the Dirac operator $D$ via the Atiyah-Singer theorem
\cite{Atiyah:1968mp}
\beq
q=n_+-n_-
\label{asit}
\eeq
where $n_\pm$ denotes the number of zero modes of $D$ with positive or negative
chirality.

On the lattice the definition of the topological charge is not unique.
Aiming for the gluonic side of (\ref{asit}), one may choose any discretization
of $F_{\mu\nu}$ which has the correct perturbative continuum limit, and form
the so-called ``naive'' (or unrenormalized field-theoretic) charge
\beq
q_\mr{nai}[U]
={1\ovr32\pi^2}\sum_x\;\trc(\ep_{\mu\nu\si\rh}F_{\mu\nu}(x)F_{\si\rh}(x))
={1\ovr 4\pi^2}\sum_x\;\trc(F_{12}F_{34}+F_{13}F_{42}+F_{14}F_{23})
\;.
\label{defqnai}
\eeq
In general this definition does not lead to an integer%
\footnote{There are two known exceptions, the L\"uscher \cite{Luscher:1981zq}
and the Phillips-Stone \cite{Phillips:1986qd} definitions of $q_\mr{fth}[U]$.}
charge.

On the other hand, the fermionic side of (\ref{asit}) always yields an integer
answer.
This is most straightforward with the massless overlap%
\footnote{It does not matter whether one uses the Wilson operator $D_\mr{W}$ or
another doubler-free kernel \cite{DeGrand:2000tf,Bietenholz:2006ni}.}
operator
$D=\rh[1+D_{\mr{W},-\rh}(D_{\mr{W},-\rh}\dag D_{\mr{W},-\rh})^{-1/2}]$
\cite{Neuberger:1997fp} or with any other Dirac operator $D$ which satisfies
the Ginsparg-Wilson relation \cite{Ginsparg:1981bj}.
In this case the index can be written in the closed form
\cite{Hasenfratz:1998ri,Luscher:1998pq}
\beq
q_\mr{fer}[U]=-{1\ovr2\rh}\trc(\gaf D)
\;.
\label{defqfer}
\eeq
With a non-chiral Dirac operator (e.g.\ $D_\mr{W}$) explicit mode counting
prescriptions may be set up.
Below, the only point which matters is that $q_\mr{fer}[U]$ is necessarily
an integer.

The continuum topological susceptibility at zero virtuality is defined as
\beq
\ch_\mr{top}=\lim_{V\to\infty}{\<q^2\>\ovr V}
\label{defchicon}
\eeq
and this shows that a finite volume is mandatory for the definition.
Still, contributions to $\<q^2\>$ which grow less than linearly in the
4D volume $V$ create finite volume effects in $\ch_\mr{top}$.

On the lattice the details of the topological charge definition reflect
themselves in the precise form of the latticized version of (\ref{defchicon})
\cite{Gockeler:1989qg}.
In case one starts with the gluonic definition (\ref{defqnai}), the traditional
approach has been to form the bare susceptibility $\<q_\mr{nai}^2\>/V$ which is
then subject to both additive and multiplicative renormalization
\cite{Campostrini:1989dh,DiGiacomo:1991ba,Alles:1997nu}.
On the other hand, starting from the fermionic definition (\ref{defqfer}), one
just forms $\<q_\mr{fer}^2\>/V$, since $q_\mr{fer}$ is already a renormalized
charge.
Likewise, if we first compute a renormalized (integer) field-theoretic charge,
the susceptibility based on it will not require any further renormalization%
\footnote{Note that the round-to-integer operation (\ref{defqren}) brings in a
global element, and our $q_\mr{ren}$ is \emph{not} given by the integral over a
local charge density. Accordingly, there is no conflict with the result by
Stamatescu and Seiler that in general $q(x)$ mixes with the identity and the
correlator $\int\!q(x)q(0)\,dx$ has a contact term \cite{Seiler:1987ig}. Any
overlap based charge definition sticks out in the sense that it is an integral
of a local density \emph{and} avoids the mixing with the identity
\cite{Chandrasekharan:1998wg,Niedermayer:1998bi,Giusti:2004qd}. The argument
for the absence of additive mass renormalization is then standard -- in the
zero-charge sector the susceptibility vanishes exactly. Note that it is
essential to define the sectors with the same operator that is used in the
$\ch_\mr{top}$ estimator -- if one uses one overlap charge to define the sector
and another overlap charge to measure $\<q^2\>$, then even the overlap based
susceptibility has an additive renormalization.}.

\begin{figure}
\epsfig{file=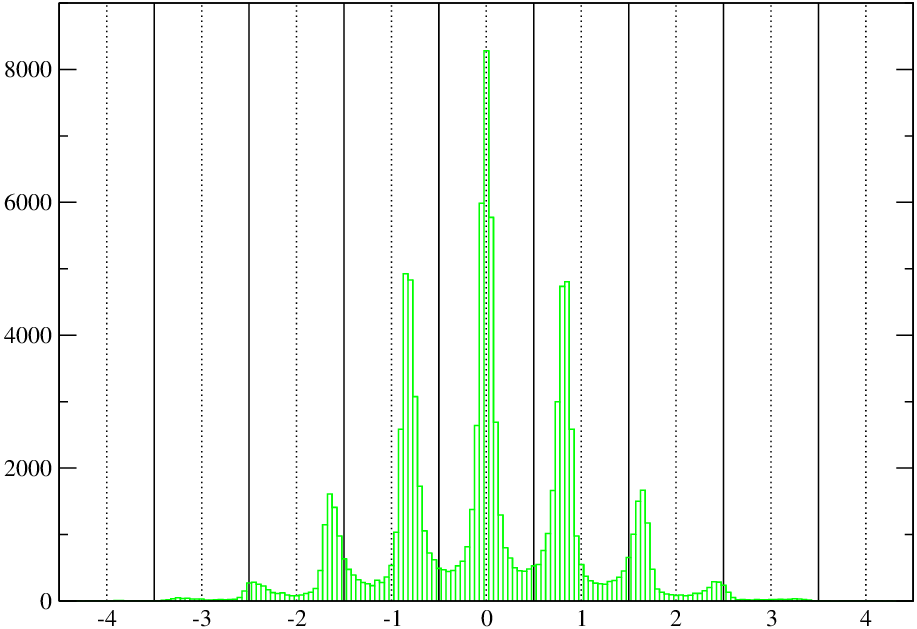,width=8.4cm}
\epsfig{file=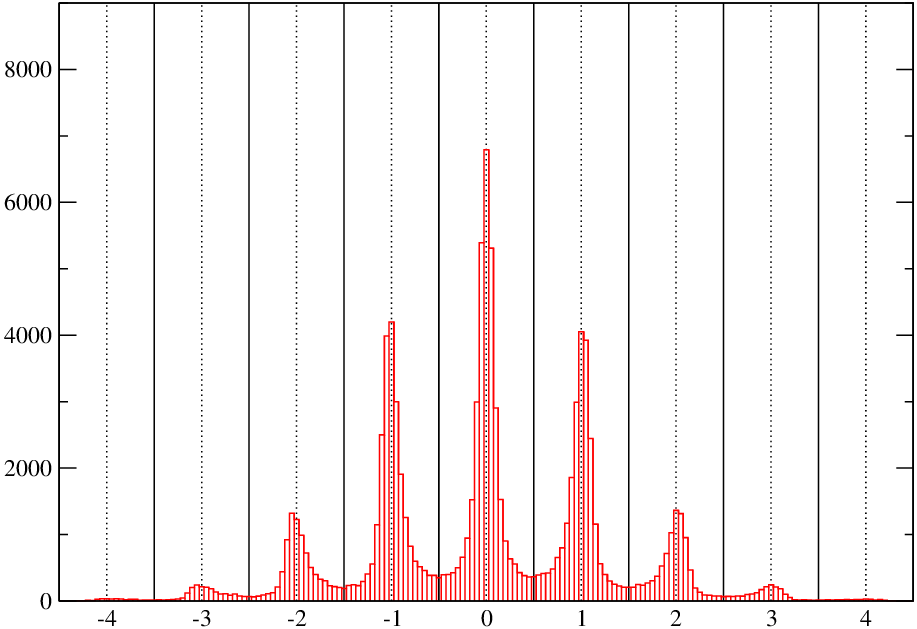,width=8.4cm}
\vspace{-2mm}
\caption{Histogram of the topological charge $q_\mr{nai}$ with 3 HYP steps
for the $12^4$ lattices at $\be=6.0$, before and after rescaling with the
renormalization factors defined in (\ref{defz}).}
\label{fig:dist}
\end{figure}

In this paper we investigate the utility of such a renormalized gluonic
charge definition for a precision measurement of the topological susceptibility
in the pure $SU(3)$ gauge theory.
We start from the standard ``clover-leaf'' definition of $F_{\mu\nu}(x)$ (it
uses the average of the antihermitean part of the plaquette $U_{\mu\nu}$ in
$x,x\!-\!\hat{\mu},x\!-\!\hat{\nu},x\!-\!\hat{\mu}\!-\!\hat{\nu}$) based on HYP
smeared \cite{Hasenfratz:2001hp} gauge links.
Plugging this into (\ref{defqnai}) we have the bare charge $q_\mr{nai}[U]$
which is a real number.
One of our $q_\mr{nai}$ distributions ($\be\!=\!6.0, 12^4$, 3\,HYP steps) is
shown in the left panel of Fig.\,\ref{fig:dist}.
Thanks to CP symmetry only a multiplicative renormalization applies, and we opt
for a non-perturbatively defined $Z$-factor.
Still, there are various possibilities, and we choose a strategy which makes
use of the fact that on fine enough lattices the overall distribution of
$q_\mr{nai}$ tends to cluster near integer values (cf.\ Fig.\,\ref{fig:dist}).
We find $Z$ as the solution%
\footnote{The restriction $Z>1$ is a technical aspect of the minimization
procedure to avoid the global minimum $\ch^2=0$ at $Z=0$ \cite{Durr:2004xu}.
In perturbation theory one finds $Z=1+\mr{const}\,g_0^2$ with $\mr{const}>0$
\cite{Christou:1995zn}.}
of
\beq
\min_{Z>1}(\ch^2)
\qquad\mr{where}\qquad
\ch^2=\sum_{U}\Big(Z\,q_\mr{nai}[U]-\mr{round}(Z\,q_\mr{nai}[U])\Big)^2
\label{defz}
\eeq
and use it to define, via rounding to the nearest integer, the renormalized
field-theoretic charge
\begin{equation}
q_\mr{ren}[U]=\mr{round}(Z\,q_\mr{nai}[U])
\label{defqren}
\eeq
which, by construction, is an integer.
This charge definition has already been used in \cite{Durr:2004xu}.

The alert reader might be surprised by our frequent use of expressions like
``we choose'' or ``we opt for'' in this passage.
Indeed, there is a huge amount of freedom in how one attributes
an integer charge to a lattice configuration.
However, according to the standard scaling hypothesis by Symanzik this
ambiguity reflects itself in different $O(a^2)$ cut-off effects of observables
built from the topological charge.
It has been checked that the ``disagreement rate'' between any pair of charge
definitions quickly vanishes with $\be\!\to\!\infty$ \cite{Gattringer:1997qc}
and that the gluonic or fermionic charge definition yields the same continuum
limit for the topological susceptibility
\cite{Alles:1997nu,Cundy:2002hv,DelDebbio:2003rn}.
Clearly, this is not a mathematical proof, but it is worth emphasizing that
this is exactly the behavior that one expects to see, if one is in the Symanzik
scaling regime.


\section{Lattice simulations}


Our goal is to perform a series of simulations in a fixed physical volume that
will allow us to determine the topological charge distribution in the continuum
limit (in that volume).
This will be complemented by a second series of simulations (at a fixed lattice
spacing) to assess possible finite volume effects.

\begin{table}[!b]
\centering
\begin{tabular}{|l|cccccc|}
\hline
$\be$                       &  5.8980 &  6.0000 &  6.0938 &  6.1802 &  6.2602 &  6.3344 \\
$L/a$                       &    10   &    12   &    14   &    16   &    18   &    20   \\
\hline
$n_\mr{sepa}$               &    10   &    10   &    10   &    50   &    50   &   100   \\
$n_\mr{conf}$               &  100010 &  101600 &  103705 &  101710 &  112222 &  105314 \\
$Z$                         &  1.2902 &  1.2353 &  1.1993 &  1.1742 &  1.1570 &  1.1430 \\
$\ta_\mr{int}(q)$           &   0.93  &   2.18  &   4.83  &   2.27  &   4.65  &   4.61  \\
$\ta_\mr{int}(q^2)$         &   0.61  &   1.09  &   2.44  &   1.19  &   2.45  &   2.52  \\
$\ta_\mr{int}(\mr{sign}(q))$&   0.89  &   2.05  &   4.42  &   2.10  &   4.25  &   4.27  \\
$|\<q\>|$                   &0.004(5) &0.007(8) &0.007(11)&0.007(8) &0.021(10)&0.010(10)\\
$\<q^2\>$                   &1.695(9) &1.592(11)&1.490(15)&1.465(10)&1.427(14)&1.381(14)\\
$\<q^4\>/\<q^2\>^2-3$       &0.214(21)&0.238(24)&0.244(27)&0.227(21)&0.241(30)&0.204(32)\\
$\<q^4\>/\<q^2\>-3\<q^2\>$  &0.363(36)&0.378(38)&0.363(40)&0.333(32)&0.344(43)&0.282(44)\\
$\<q^4\>-3\<q^2\>^2$        &0.616(61)&0.602(62)&0.541(60)&0.487(46)&0.490(63)&0.389(61)\\
$\<q^2\>_{q\neq0}$          &1.702(12)&1.610(13)&1.526(22)&1.486(13)&1.461(21)&1.400(19)\\
\hline
\end{tabular}
\caption{Summary of the scaling series of runs. Unless stated otherwise, $q$ is
the renormalized (i.e.\ integer) gluonic topological charge $q_\mr{ren}$ with
3\,HYP steps. The subscript ``$q\!\neq\!0$'' means that the distribution of
$|q|$ with $q\!\neq\!0$ has been fitted to a half-Gaussian. Statistical errors
have been estimated via a jackknife with the blocklength set to
$\mr{round}(10\ta_\mr{int}(q))$.}
\label{tab:1}
\vspace{4mm}
\begin{tabular}{|l|ccccc|}
\hline
$\be$                       &  6.0000 &  6.0000 &  6.0000 &  6.0000 &  6.0000 \\
$L/a$                       &    10   &    12   &    14   &    16   &    18   \\
\hline
$n_\mr{sepa}$               &    10   &    10   &    10   &    10   &    10   \\
$n_\mr{conf}$               &  103600 &  101600 &   99000 &  100000 &  105000 \\
$Z$                         & [1.2441]&  1.2353 & [1.2333]& [1.2332]& [1.2237]\\
$\ta_\mr{int}(q)$           &   1.80  &   2.18  &  2.25   &   2.09  &   2.13  \\
$\ta_\mr{int}(q^2)$         &   1.34  &   1.09  &  1.10   &   1.11  &   1.13  \\
$\ta_\mr{int}(\mr{sign}(q))$&   1.74  &   2.05  &  1.96   &   1.74  &   1.71  \\
$|\<q\>|$                   &0.004(4) &0.007(8) &0.003(11)&0.020(14)&0.023(17)\\
$\<q^2\>$                   &0.586(5) &1.592(11)&3.012(20)&5.198(34)&8.233(53)\\
$\<q^4\>/\<q^2\>^2-3$       &1.277(41)&0.238(24)&0.095(20)&0.065(19)&0.008(16)\\
$\<q^4\>/\<q^2\>-3\<q^2\>$  &0.748(24)&0.378(38)&0.287(59)&0.338(97)&0.069(139)\\
$\<q^4\>-3\<q^2\>^2$        &0.439(15)&0.602(62)&0.86 (18)&1.76 (50)&0.57 (114)\\
$\<q^2\>_{q\neq0}$          &0.780(10)&1.610(13)&3.009(26)&5.202(51)&8.242(54)\\
\hline
\end{tabular}
\caption{Summary of the volume extrapolation series of runs. Throughout, the
charge renormalization factors of the ($6.0,12^4$) lattices have been used. For
details see caption of Tab.\,1.}
\label{tab:2}
\vspace{-4mm}
\end{table}

We use the MILC code \cite{milccode} to produce the $SU(3)$ gauge ensembles.
We choose the Wilson gauge action and run parameters as detailed in
Tab.\,\ref{tab:1} (for the scaling series) and Tab.\,\ref{tab:2} (for the
volume extrapolation series).
The scale is set via the Sommer radius $r_0$ \cite{Sommer:1993ce}, based on a
new parameterization of data from \cite{Necco:2001xg} as specified in the
appendix.
The physical value is a separate topic that will be discussed along with the
presentation of the final result.

We start with a closer look at Tab.\,\ref{tab:1}.
Throughout, our runs are designed to yield $\sim\!10^5$ measurements.
Anticipating that the autocorrelation time of the topological charge rapidly
grows with $\be$, we have increased the number of updates, $n_\mr{sepa}$,
between adjacent measurements.
Still, it turns out that the measured integrated autocorrelation times are
somewhat larger on the finer lattices.
The renormalization factor $Z$ for $q_\mr{nai}$ with 3\,HYP steps, as defined
in (\ref{defz}), seems to go monotonically towards 1 with increasing $\be$, as
expected.
Based on it we determine $q\!=\!q_\mr{ren}$ defined in (\ref{defqren}).
As a first check, we measure $|\<q\>|$, and this moment is consistent with zero
on all lattices.
The observable of interest, $\<q^2\>$, is measured with 1\% statistical
accuracy (or better), throughout.
We have also checked that using a 2\,HYP or 4\,HYP charge would change the
suscpetibility by an amount which is an order of magnitude smaller than the
statistical error of the $\<q^2\>$ given.
Our data are precise enough to evaluate the fourth moment of the distribution,
and the kurtosis $\<q^4\>/\<q^2\>^2-3$ turns out to deviate from zero, for most
lattices in the scaling series, by about $10\si$.
It seems independent of the lattice spacing and it is thus natural to ask
whether a non-vanishing $\<q^4\>/\<q^2\>^2-3$ is a finite volume effect.

This brings us to discuss the second series of runs, as detailed in
Tab.\,\ref{tab:2}.
Potential finite volume effects are commonly attributed to infrared physics; we
thus restrict ourselves to a single coupling, $\be\!=\!6.0$.
Again, we aim for $O(10^5)$ measurements per run.
For larger volumes the ``crest and valley'' structure in Fig.\,\ref{fig:dist}
becomes less pronounced%
\footnote{Here it is understood that the total statistics is kept fixed, so
that the number of configurations in a given sector drops. Via appropriately
increasing the statistics, $Z$ can be determined in a volume as large as
desired.},
and we pursue the analysis with the $Z$-factor determined on the $(6.0,12^4$)
lattices.
Now one expects the autocorrelation times to be independent of the volume.
This happens to be true, except in the $L/a\!=\!10$ case, where we find severe
finite volume effects in $\<q^2\>$, too (see below).
Again $|\<q\>|$ is basically consistent with zero.
The second moment $\<q^2\>$ is measured, as before, with 1\% statistical
accuracy (or better) on all lattices.
The main point of this series is clear evidence that the kurtosis
$\<q^4\>/\<q^2\>^2-3$, which quantifies the deviation from a Gaussian
distribution, tends to zero with $V\!\to\!\infty$.

\begin{figure}[!b]
\epsfig{file=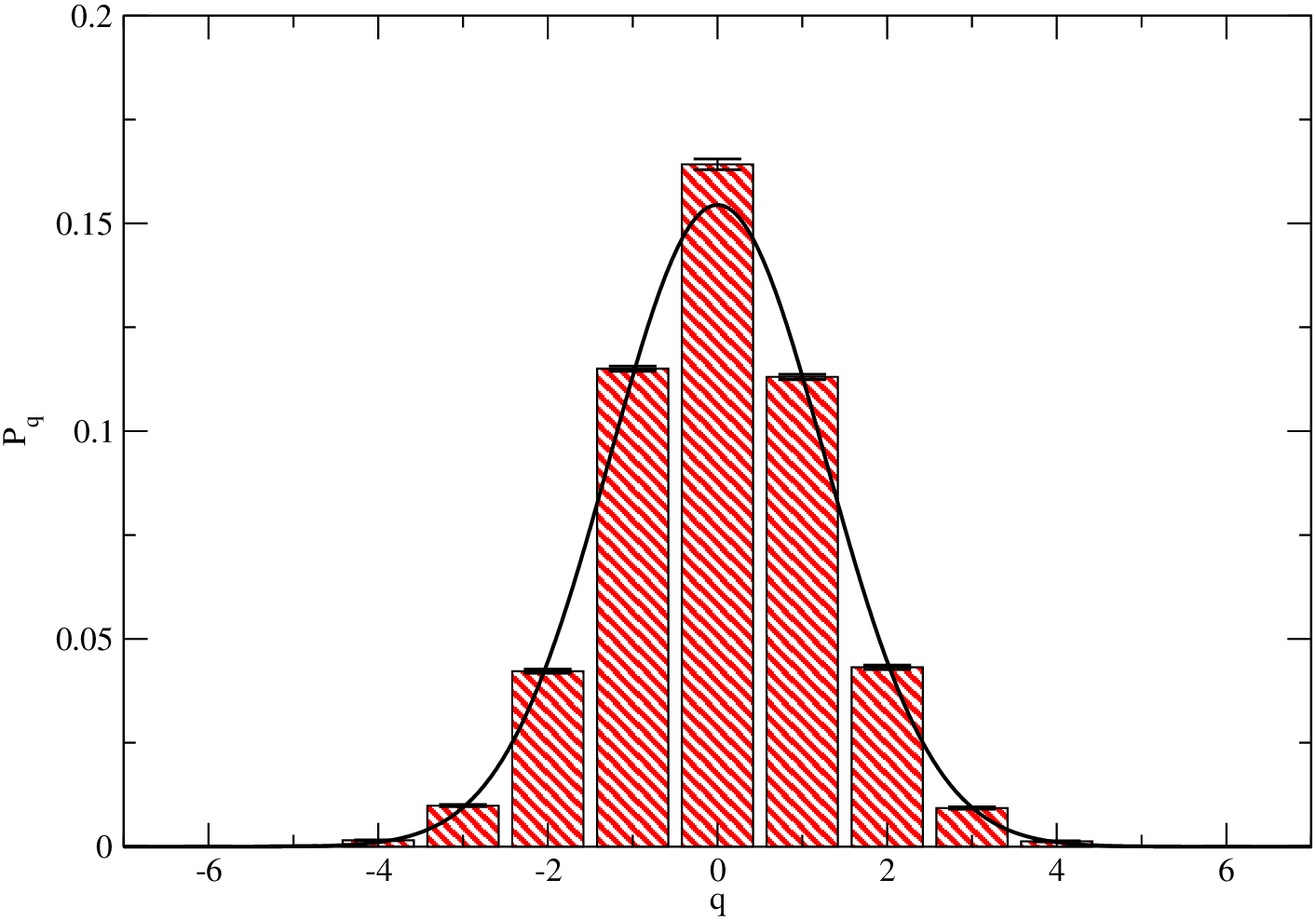,width=8.4cm}
\epsfig{file=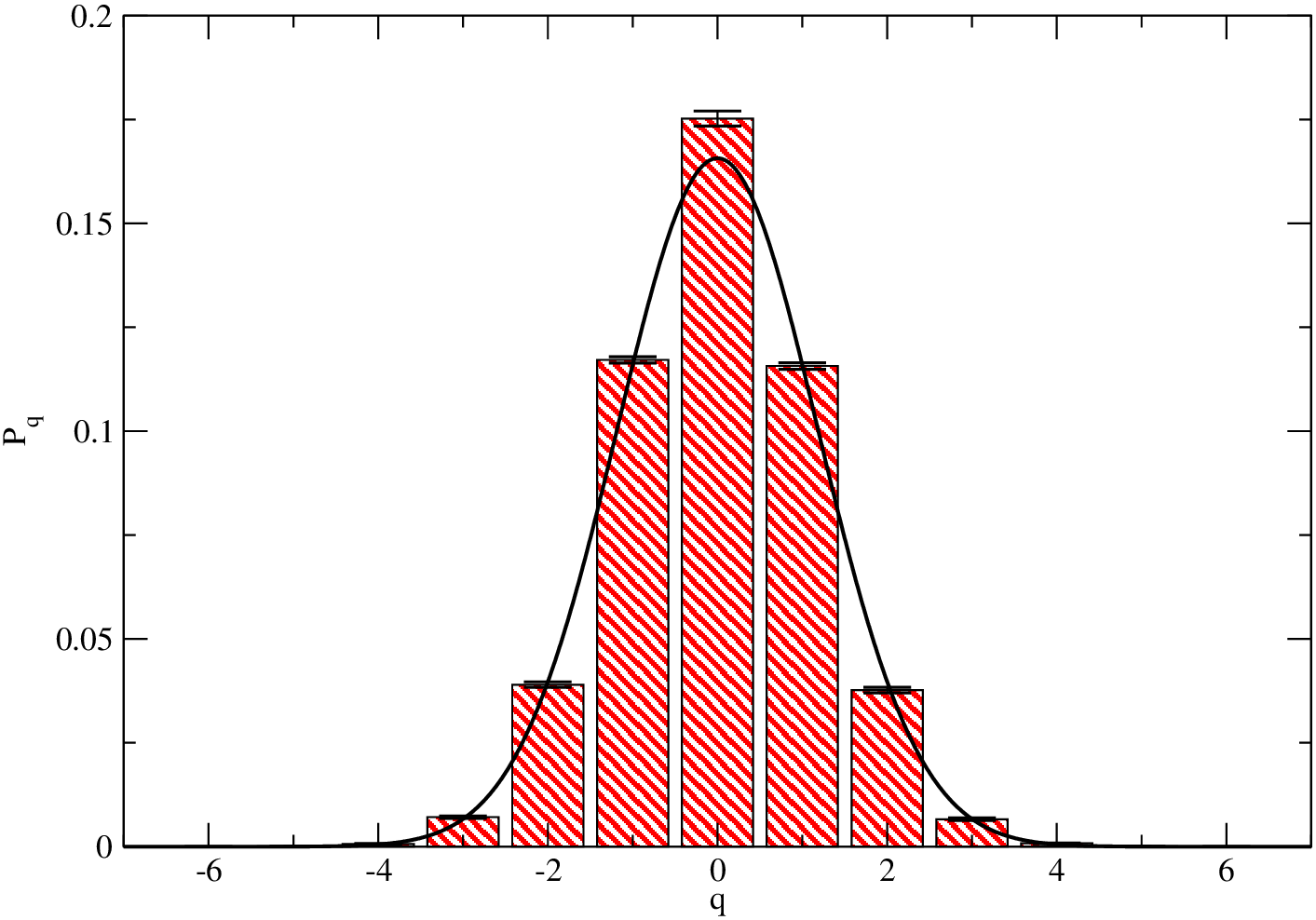,width=8.4cm}
\vspace{-2mm}
\caption{Distribution of $q_\mr{ren}$ for the $(6.0,12^4)$ [left] and
$(6.3344,20^4)$ lattices [right]. The fit to a Gaussian form excludes the
$q\!=\!0$ sector; the excess gives rise to the kurtosis
$\<q^4\>/\<q^2\>^2\!-\!3$ reported in Tab.\,1, an effect that goes away in the
infinite volume limit, as evident from Tab.\,2.}
\label{fig:hist}
\end{figure}

This observation lets us plot the charge histograms for two lattices from the
scaling series in Fig.\,\ref{fig:hist}.
It turns out that the relative weight of the toplogically non-trivial sectors
is almost consistent with a Gaussian form, while the $q\!=\!0$ sector shows a
clear excess.
It is thus tempting to define a new $\<q^2\>_{q\neq0}$ as the width of this
Gaussian, where the fit takes only the nontrivial sectors into account.
In fact, due to CP symmetry the distribution is even, and we produce a
histogram of $|q|$.
The $|q|\!\geq\!1$ sectors are then fitted to a half-Gaussian.
This procedure stabilizes the contribution from the tails (for this reason a
similar ``improved estimator'' was used in \cite{DelDebbio:2004ns}), but it
also introduces a model dependence.

\section{Continuum and infinite volume extrapolations}


\begin{table}
\centering
\begin{tabular}{|l|ccccc|}
\hline
$\be$                          & 6.0000 & 6.0938 & 6.1802 & 6.2602 & 6.3344 \\
$L/a$                          &   12   &   14   &   16   &   18   &   20   \\
\hline
$\<q^2\>(r_0/L)^4$             &6330(60)&5926(71)&5826(57)&5673(67)&5490(66)\\
$\<q^2\>_{q\neq0}(r_0/L)^4$    &6401(69)&6066(95)&5907(66)&5808(92)&5568(84)\\
\hline
\end{tabular}
\caption{Summary of the data from the scaling series of runs, in units of
$10^{-5}$. Here, an extra 0.68\% error has been added w.r.t.\ what is reported
in Tab.\,1.}
\label{tab:3}
\vspace{4mm}
\begin{tabular}{|l|ccccc|}
\hline
$L/a\;\;(\be\!=\!6.0)$         &   10   &   12   &   14   &   16   &   18   \\
\hline
$\<q^2\>(r_0/L)^4$             &4834(42)&6330(42)&6464(43)&6540(43)&6468(42)\\
$\<q^2\>_{q\neq0}(r_0/L)^4$    &6428(79)&6401(53)&6458(56)&6545(64)&6474(42)\\
\hline
\end{tabular}
\caption{Summary of the data from the infinite volume extrapolation series, in
units of $10^{-5}$. Here, no extra error has been added w.r.t.\ what is
reported in Tab.\,2.}
\label{tab:4}
\vspace{-2mm}
\end{table}

For the continuum and infinite volume extrapolation of the quenched topological
susceptibility we convert our data to physical units by multiplying them with
the appropriate power of $r_0/a$.
In the scaling series the uncertainty of $L/r_0$ reflects itself independently
for each datapoint.
Assuming that $L/r_0$ is known (at each $\be$) with 0.17\% precision (see
Appendix) an extra 0.68\% error needs to be included in Tab.\,\ref{tab:3}.
By contrast, in the volume extrapolation series the uncertainty of $L/r_0$ is
correlated, and no extra error has been included in Tab.\,\ref{tab:4}.

\begin{figure}
\epsfig{file=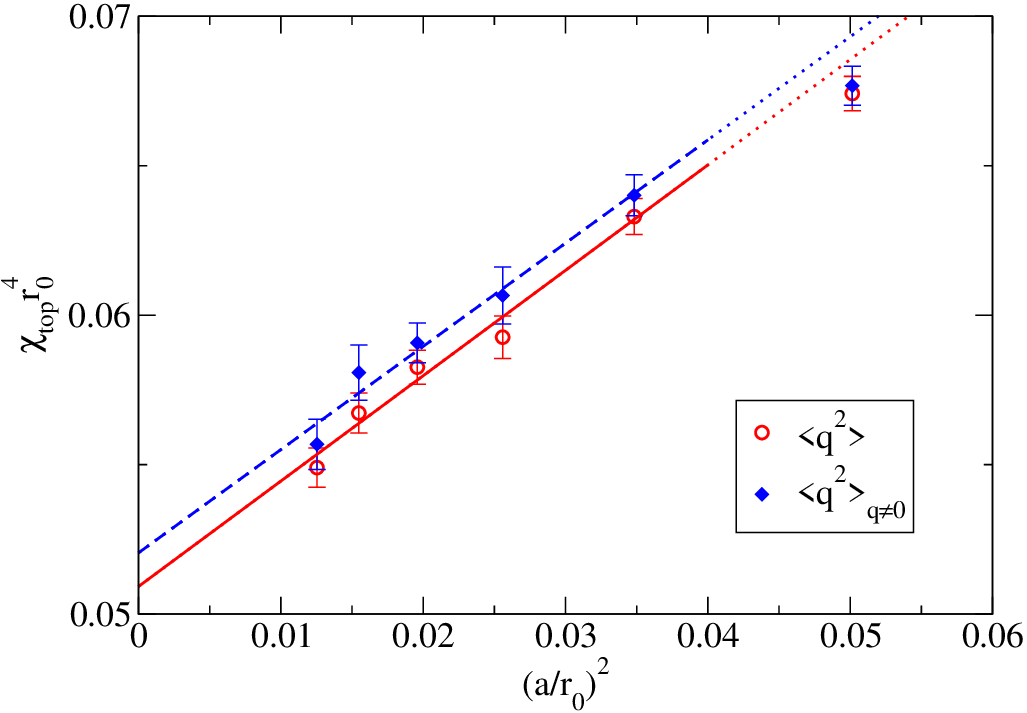,width=17cm}
\vspace{-6mm}
\caption{Continuum extrapolation of $\ch_\mr{top}r_0^4$ in a fixed physical
volume, $V=(2.2394r_0)^4$.}
\label{fig:contin}
\vspace{8mm}
\epsfig{file=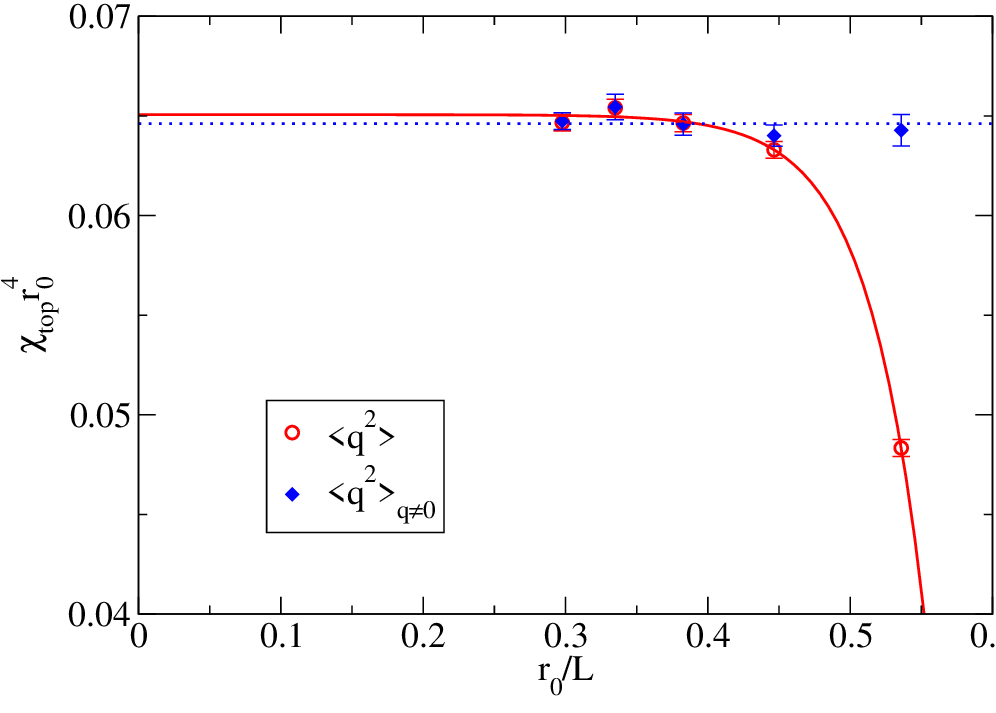,width=8.4cm}
\epsfig{file=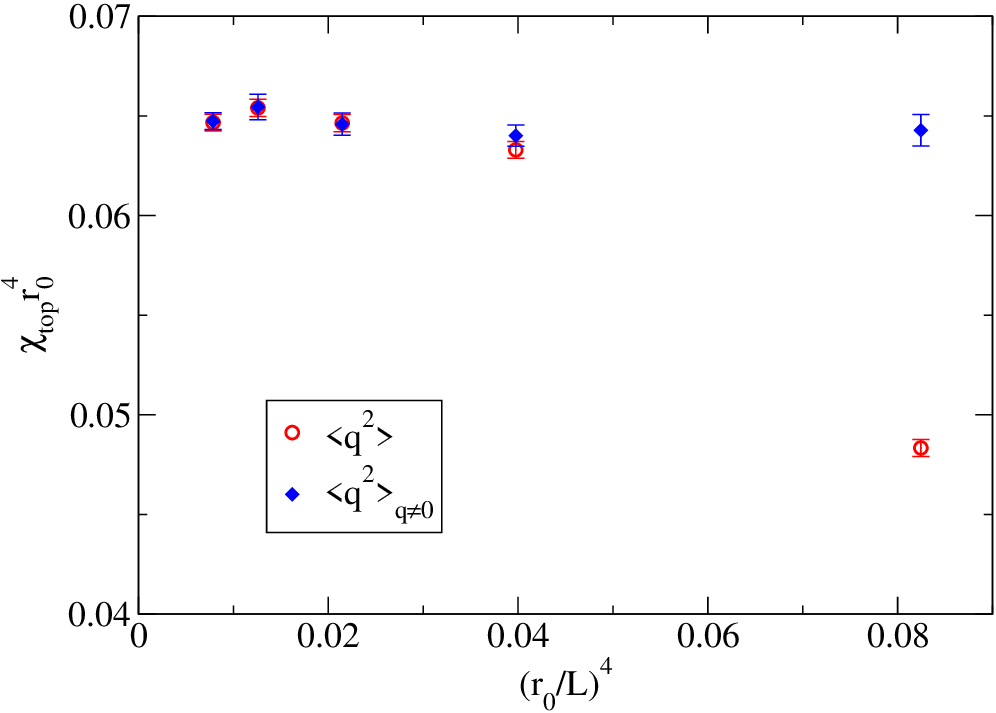,width=8.4cm}
\vspace{-2mm}
\caption{Infinite volume extrapolation of $\ch_\mr{top}r_0^4$, versus
$r_0/L$ [cf.\ (\ref{infvol1})] and $(r_0/L)^4$ [cf.\ (\ref{infvol2})].}
\label{fig:infvol}
\end{figure}

\begin{figure}
\epsfig{file=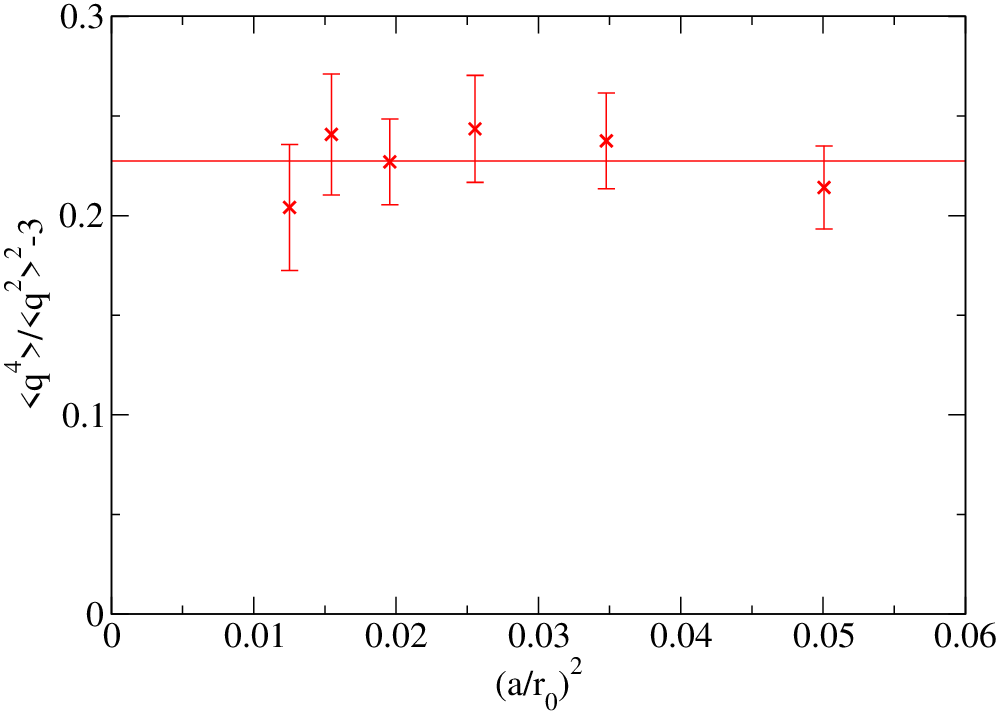,width=17cm}
\vspace{-6mm}
\caption{Continuum extrapolation of $\<q^4\>/\<q^2\>^2-3$ in a fixed physical
volume, $V=(2.2394r_0)^4$.}
\label{fig:kurcon}
\vspace{8mm}
\epsfig{file=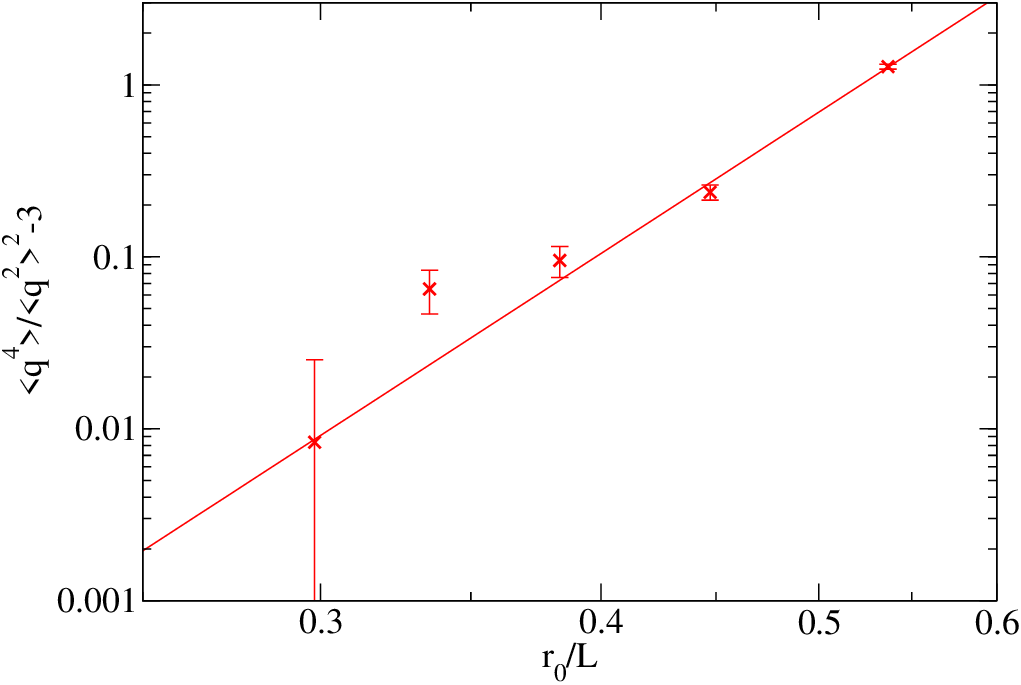,height=5.9cm}
\epsfig{file=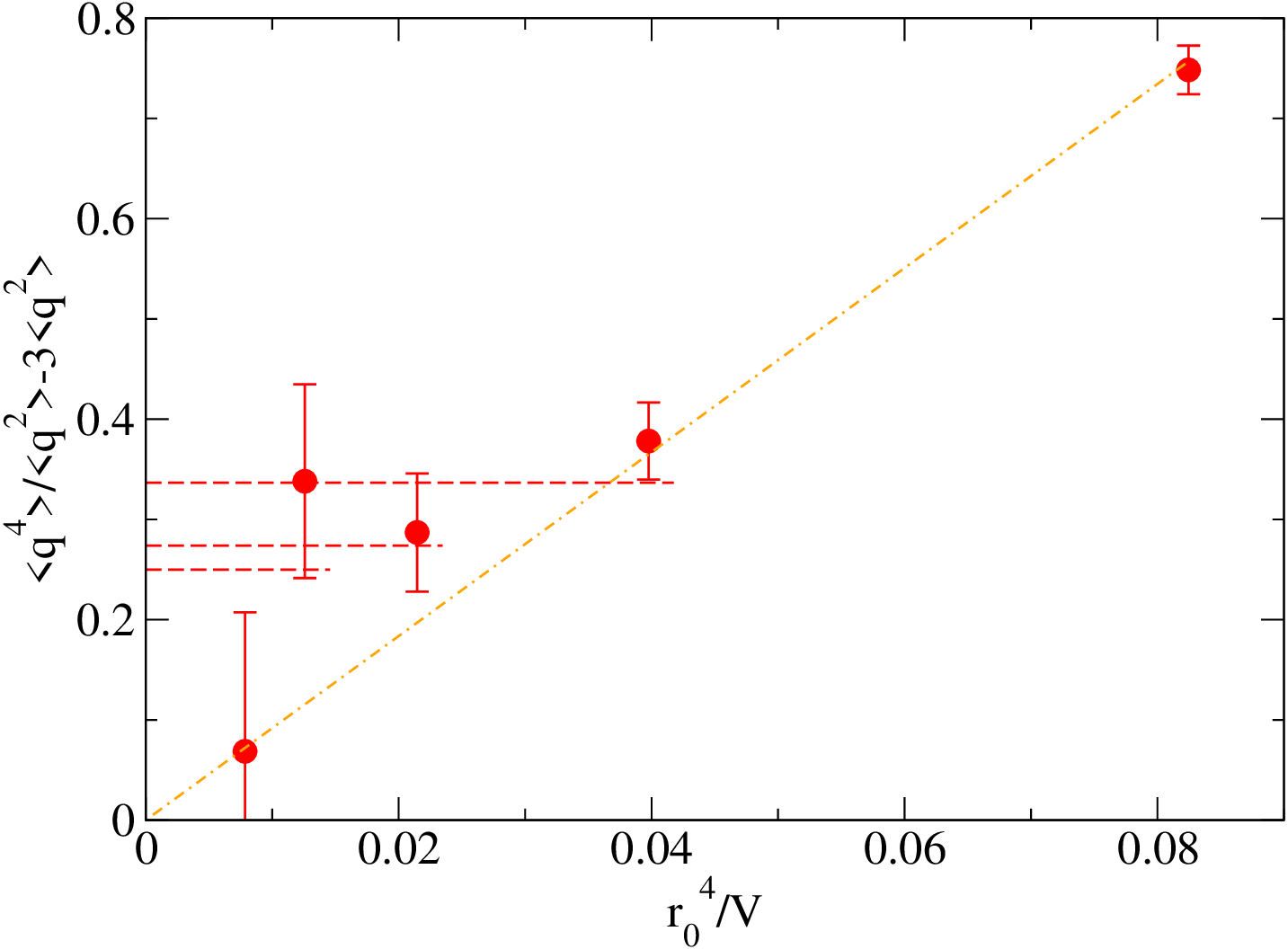,height=5.9cm}
\vspace{-2mm}
\caption{Left: Infinite volume extrapolation of the kurtosis
$\<q^4\>/\<q^2\>^2-3$ in log-log form; this quantity vanishes with
$V\!\to\!\infty$. Right: The reduced moment ratio $\<q^4\>/\<q^2\>-3\<q^2\>$
versus $r_0^4/V$. This plot includes constant fits to the last 2, 3 or 4
datapoints and a linear fit to all 5 points. We cannot discriminate a finite
from a vanishing limit of this quantity with $V\!\to\!\infty$.}
\label{fig:kurvol}
\end{figure}

For the continuum extrapolation in a fixed physical volume we plot our data
against $(a/r_0)^2$.
This power of the lattice spacing is motivated by a Symanzik analysis 
(both the Wilson action and our topological charge operator contain dimension
6, but no dimension 5 operators).
In principle, the uncertainty of $r_0/a$ also leads to a horizontal error bar,
but its effect is negligible, and we shall omit it.
The resulting fits are shown in Fig.\,\ref{fig:contin}.
The observables $\<q^2\>(r_0/L)^4$ and $\<q^2\>_{q\neq0}(r_0/L)^4$ extrapolate
in a parallel manner, giving the continuum values $0.05092(71)$ and
$0.05205(71)$, respectively.
The difference, if evaluated inside a jackknife, is significant.

This brings us to the discussion of finite volume effects.
Standard reasoning suggests that the dominant finite volume corrections in the
YM theory come from glueball states travelling around the box.
Since euclidean Green's functions fall off exponentially, this would imply
\beq
Z_\nu(L)=Z_\nu(\infty)\,\Big(1+\mr{const}\,e^{-M_G L}+...\Big)
\label{infvol1}
\eeq
in close analogy to relative finite volume effects in QCD being proportional
to $\exp(-\Mpi L)$.
Simple models for the YM vacuum (random population of unit cells by instantons
or anti-instantons) suggest no finite volume corrections for the susceptibility
and a $1/V$ type kurtosis%
\footnote{With \emph{independent} fluctuations in unit cells $V_1,...,V_N$
this is generic: $\<\nu^2\>=\<(\nu_1+...+\nu_N)^2\>=N\<\nu_1^2\>$ and
$\<\nu^4\>=\sum_i^N\<\nu_i^4\>+6\sum_{i<j}^N\<\nu_i^2\>\<\nu_j^2\>=
N\<\nu_1^4\>+3N(N-1)\<\nu_1^2\>\<\nu_2^2\>$, where we use that all odd moments
vanish.}.
As a result, one may consider the alternative form of the partition function
\beq
Z_\nu(L)=Z_\nu(\infty)\,\Big(1+\mr{const}/L^4+...\Big)
\;.
\label{infvol2}
\eeq
The data of Tab.\,\ref{tab:4} are shown in Fig.\,\ref{fig:infvol}.
If we assume exponential finite-volume effects, the data for the first
estimator may be fitted all the way out to $L/a\!=\!10$.
The difference of the fit curve at $L/a\!=\!12$ and at $L/a\!=\!\infty$
suggests that the continuum result for $\<q^2\>(r_0/L)^4$ is suppressed by the
amount $0.00175$, and this gives a corrected $\<q^2\>(r_0/L)^4\!=\!0.05267(71)$.
With the other estimator the fitted value for ``const'' in (\ref{infvol1}) is
consistent with zero, and we stay with
$\<q^2\>_{q\neq0}(r_0/L)^4\!=\!0.05205(71)$, as quoted above.
If we assume $1/V$-type effects, finite-volume corrections are hard to quantify
(at least the volume-scaling regime with the first estimator is smaller).
Therefore, we decide to stay with the extrapolation via (\ref{infvol1}).
Taking the average of both estimators but keeping the full difference as a
theoretical uncertainty, we consider
\beq
\ch_\mr{top}r_0^4=0.05236(71)(62)
\eeq
our final result, where the first error is statistical and the second
systematic.


Our data for the kurtosis $\<q^4\>/\<q^2\>^2-3$ scale perfectly
(Fig.\,\ref{fig:kurcon}), and the infinite volume extrapolation is consistent
with both an exponential and a power law behavior.
We find that in the YM theory a non-zero kurtosis is a pure finite volume
effect and has nothing to do with the discretization  (Fig.\,\ref{fig:kurvol}).
Since our ``independent cell'' model of footnote 7 and large $\Nc$ arguments
suggest that $\<q^4\>/\<q^2\>-3\<q^2\>$ has a finite limit with
$V\!\to\!\infty$, we plot this quantity versus $1/V$ in Fig.\,\ref{fig:kurvol},
too.
It turns out that our data are consistent with the value $0.276(84)$ given in
\cite{DelDebbio:2002xa}, but we cannot rule out the possibility that the
infinite volume limit might actually vanish.


\section{Conversion to physical units}


Having an unambiguous result for the topological susceptibility in the combined
continuum and infinite volume limit in units of the Sommer scale $r_0^{-1}$
\cite{Sommer:1993ce}, we are left with the question which value [in MeV] the
latter should be identified with.

In QCD $r_0$ is a well-defined quantity.
In other words, one could measure $aM_p$, $a\Mpi$ and $aM_K$ in a $2\!+\!1$
flavor simulations in which their ratios are always adjusted to the respective
experimental ratios.
By considering the continuum limit of $r_0M_p$, one would have the correct
physical value of $r_0$, and there is nothing wrong with using this value
also in pure YM theory.
The original estimate $r_0\!=\!0.5\fm$ \cite{Sommer:1993ce} for the outcome
of this procedure has been superseded, more recently, by values like
$r_0\!=\!0.467(6)\fm$ \cite{Khan:2006de,Aubin:2004wf}.
Alternatively, one could set the scale in the quenched theory via
$r_0f_K\!=\!0.4146(94)$ \cite{Garden:1999fg}, using
$f_K\!=\!160(2)\MeV$ from experiment.
This has the same effect as attributing $r_0\!=\!0.512(12)\fm$.
To encompass this spread we set the scale via $r_0\!=\!0.49\fm$ and add a 4\%
error to reflect the intrinsic scale ambiguity in pure YM theory.

With this choice for $r_0$ our value $\ch_\mr{top}r_0^4$, as quoted in the
previous section, amounts to
\beq
\ch_\mr{top}^{1/4}=193(1)(8)\MeV
\label{final}
\eeq
where the first error bar contains all statistical and systematic uncertainties
of our calculation, and the second one reflects the scale setting ambiguity in
a theory which is not full QCD.


\section{Summary}


In this paper we have performed a precision study of the topological
susceptibility in pure $SU(3)$ Yang Mills theory in the combined continuum and
infinite volume limit using a field-theoretic definition of the topological
charge.

\begin{table}
\centering
\begin{tabular}{|c|lcc|}
\hline
&$\;\ch_\mr{top}r_0^4$&$\ch_\mr{top}^{1/4}$ [MeV]&$\ch_\mr{top}^{1/4}$ [MeV]\\
\hline
Ref.\,\cite{Lucini:2001ej}    (2001) &0.072(7)  & 209(5)(8) &    ---    \\
Ref.\,\cite{DelDebbio:2002xa} (2002) &0.057(3)  & 197(3)(8) & 180(2)(8) \\
Ref.\,\cite{DelDebbio:2003rn} (2003) &0.055(10) & 195(9)(8) & 188(12)(8)\\
Ref.\,\cite{Giusti:2003gf}    (2003) &0.059(5)  & 198(4)(8) &    ---    \\
Ref.\,\cite{DelDebbio:2004ns} (2004) &0.059(3)  & 198(3)(8) & 191(5)(8) \\
this work                            &0.0524(9) &\multicolumn{2}{c|}{193(1)(8)}\\
\hline
\end{tabular}
\caption{Summary of the quenched topological susceptibility (in the continuum)
as determined in some recent studies. The results
$\ch_\mr{top}/\si^2\!=\!0.0355(33)$ \cite{Lucini:2001ej} and
$\ch_\mr{top}/\si^2\!=\!0.0282(12)$ \cite{DelDebbio:2002xa} have been converted
to $r_0^{-4}$ units by means of $\si^{1/2}r_0=1.193(10)$ 
\cite{Niedermayer:2000yx}. In the second column the conversion to MeV has been
done with our choice $r_0\!=\!0.49\fm$, while the third one contains the values
given in the respective papers, but with our standard scale setting error
throughout.}
\label{tab:5}
\end{table}

Comparing our final result to other recent%
\footnote{We like to draw the readers attention to two studies
\cite{Gattringer:2002mr,Hasenfratz:2002rp} with a fermionic charge where no
continuum limit has been taken. Note that the charges $Q_I$ in
\cite{Lucini:2001ej} and $Q_g$ in Fig.5 of \cite{DelDebbio:2003rn} are similar
in spirit to our $q_\mr{ren}$.}
determinations \cite{Lucini:2001ej,DelDebbio:2002xa,DelDebbio:2003rn,
Giusti:2003gf,DelDebbio:2004ns} (see Tab.\,\ref{tab:5} for details) we see a
downward trend over time.
Our result in $r_0^{-4}$ units is substantially more precise, since we work
much closer to the continuum (our $\be\!=\!6.0$ lattices are the coarsest ones
used in the continuum extrapolation) and due to the large statistics%
\footnote{There is a similar high-statistics study, based on an overlap charge
\cite{PoS_Lat06_058}. They have no continuum limit yet for $\ch_\mr{top}$, but
they see a deviation from zero in $\<q^4\>-3\<q^2\>^2$ or the kurtosis in a
fixed volume.}.
The conversion to $\MeV$ is, of course, limited by the basic scale setting
ambiguity in the quenched theory.
In any case our final result (\ref{final}) supports the Witten Veneziano
scenario for the origin of the $\et'$ mass.

\bigskip

{\bf Acknowledgments}:
We thank Ferenc Niedermayer for useful discussions.
Computations were carried out on ALiCEnext and a Pentium4 cluster at the
University of Wuppertal.
This work was supported by the Swiss NSF and under the Hungarian grant
OTKA-AT049652.


\section*{Appendix}


In Ref.\,\cite{Necco:2001xg} Necco and Sommer use a polynomial approximation
for $\log(r_0/a)$.
Specifically
\beq
\log(r_0/a)\Big|_\mr{NS}=c_0+c_1(\be-6)+c_2(\be-6)^2+c_3(\be-6)^3
\label{r0_neso}
\eeq
with $c_0\!=\!1.6804,c_1\!=\!1.7331,c_2\!=\!-0.7849,c_3\!=\!0.4428$ is
suggested in order to generate matched lattices for $5.7\leq\be\leq6.92$,
i.e.\ combinations of $(\be,L/a)$ with fixed $L/r_0$.
We have repeated the analysis with their data \cite{Necco:2001xg} and find
essentially the same $c_{0...3}$, with $\ch^2/\mr{d.o.f.}=1.39$.

Here, we wish to explore a rational approximation of $\log(r_0/a)$ which has
the merit of being consistent with perturbation theory.
In the weak coupling regime $a\!\propto\!\exp(-1/[2\be_0g_0^2])$ with the
universal coefficient $\be_0\!=\!1/(4\pi)^2\cdot[11\Nc/3-2\Nf/3]$.
In $SU(3)$ gluodynamics one has thus $a\!\propto\!\exp(-16\pi^2/132\cdot\be)$
or $\log(r_0/a)=4\pi^2/33\cdot\be+...$ and we employ the ansatz
\beq
\log(r_0/a)\Big|_\mr{new}=
{4\pi^2\ovr33}\,\be\cdot{1+d_1/\be+d_2/\be^2 \ovr 1+d_3/\be+d_4/\be^2}
\label{r0_new}
\eeq
where this constraint is built in.
Using again their data we find that the best fit is given through
$d_1\!=\!-8.2384,d_2\!=\!15.310,d_3\!=\!-2.7395,d_4\!=\!-11.526$
with $\ch^2/\mr{d.o.f.}=0.91$.

\begin{figure}
\centering
\epsfig{file=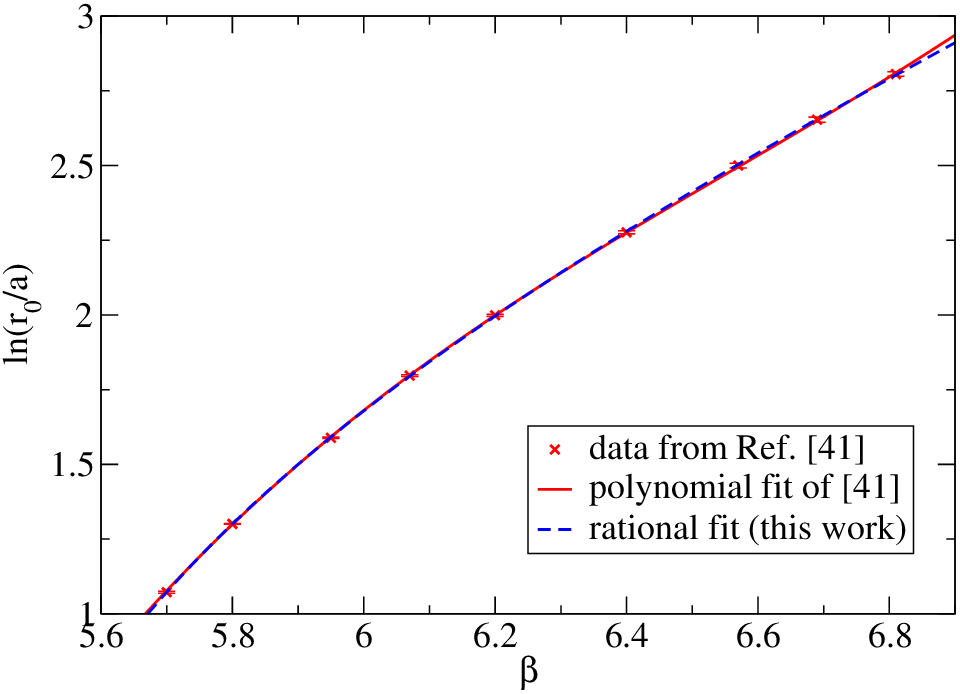,angle=0,width=15cm}
\vspace{-2mm}
\caption{Old polynomial $\ch^2/6\!=\!1.39$ and new rational $\ch^2/6\!=\!0.91$
fit to the data of Ref.\,\cite{Necco:2001xg}. Asymptotic freedom demands that
the curve eventually becomes linear with slope $4\pi^2/33$, and this constraint
is built into the new rational ansatz (\ref{r0_new}).}
\label{fig:interpolation}
\end{figure}

\begin{table}
\centering
\begin{tabular}{|c|cccccc|}
\hline
$L/a$   &   10   &   12   &   14   &   16   &   18   &   20   \\
\hline
$\be$   & 5.8996 & 6.0000 & 6.0926 & 6.1790 & 6.2601 & 6.3362 \\
$L/r_0$ & 2.2357 & 2.2356 & 2.2357 & 2.2357 & 2.2356 & 2.2357 \\
\hline
$\be$   & 5.8980 & 6.0000 & 6.0938 & 6.1802 & 6.2602 & 6.3344 \\
$L/r_0$ & 2.2394 & 2.2394 & 2.2394 & 2.2395 & 2.2393 & 2.2393 \\
\hline
\end{tabular}
\caption{Upper part: ($\be$, $L/a$) combinations which match ($6.0$, $12$)
[via achieving $L/r_0\!=\!2.2356$ as accurately as possible], based on the
interpolation formula (\ref{r0_neso}) of \cite{Necco:2001xg}. Lower part:
Same [via achieving $L/r_0\!=\!2.2394$ as accurately as possible], based on
the interpolation formula (\ref{r0_new}).}
\label{tab:6}
\end{table}

It is straightforward to find the $\be$-values which match a simulation with
$(\be,L/a)=(6.0,12)$.
The results are given in Tab.\,\ref{tab:6}, both for (\ref{r0_neso}) and for
(\ref{r0_new}).
It turns out that the matched $\be$-values are more or less the same with
either formula and the pertinent $L/r_0$ (and hence our estimates for the
physical box-length) differ by 0.17\% only.
This should not come as a surprise, since the two curves in
Fig.\,\ref{fig:interpolation} are rather close.
Still, since the new interpolation has a better theoretical foundation and the
$\ch^2/\mr{d.o.f.}$ is lower, we choose the new parameterization
(\ref{r0_new}).
Nonetheless, it is clear that the original parametrization (\ref{r0_neso})
remains a legitimate and good one in the interval in which it is given.



\end{document}